# Electronic circuits manifesting hyperbolic chaos and their simulation with software package Multisim

*S.P. Kuznetsov*

Kotel'nikov's Institute of Radio-Engineering and Electronics of RAS, Saratov Branch
Zelenaya 38, Saratov, 410019, Russian Federation

We consider several electronic circuits, which represent dynamical systems with hyperbolic chaotic attractors of Smale-Williams type, and demonstrate results of their simulation using the software package NI Multisim 10. The developed approach is useful as an intermediate step of constructing real electronic devices manifesting structurally stable hyperbolic chaos applicable e.g. in systems of secure communication, noise radar, for cryptographic systems and random number generators. This is also of methodological interest for training students who specialize in radio-physics and nonlinear dynamics in the design and analysis of systems with complex dynamics using examples close to practical applications.

## *Introduction*

Since classic works of Andronov and his school, rough, or structurally stable systems are considered as subject of priority research in the oscillation theory and as the most important from the practical point of view [1-3]. Following this principle and referring to applied aspects of dynamical chaos, one should prefer the class of systems with uniformly hyperbolic chaotic attractors, which are structurally stable [4-9]. In due time, when such attractors were introduced in the theory of dynamical systems, it was expected that they will relate to many physical situations. It turned out, however, that this is not the case: the attractors encountered in applications normally do not fit into the narrow frames of the early hyperbolic theory. Then, the efforts of mathematicians were redirected to development of generalizations embracing realistic examples of chaotic dynamics [10, 11], while the uniformly hyperbolic attractors were regarded only as refined images of chaos, which have no direct relation to reality been represented exclusively by artificial mathematical constructions.

One of these constructions, the Smale-Williams attractor, is explained often as follows [3, 5]. Consider a torus domain in three-dimensional state space of a discrete-time system. Imaging the torus as a plastic object, stretch it in length, squeeze in the transverse direction, and put inside the original domain in a form of the double loop. This is one step of the discrete time evolution. It is assumed that the volume of the object is reduced under the deformation (dissipativity). At each repetition of this transformation the number of turns of the object undergoes doubling, and it tends to infinity in the limit. The attracting limit set is the so-called solenoid, which possesses the Cantor-like structure in a transversal cross-section. Chaotic nature of the dynamics on the attractor takes place due to the instability of the motion in respect to variations of initial conditions; indeed, the longitudinal expansion along the filaments of the solenoid in the course of the dynamical evolution corresponds to increase of distance between any initially close representative points at each iteration step.

An alternative is a formal construction of an explicit map possessing the solenoid type attractor [7, 8]. Let $X$ and $Y$ be Cartesian coordinates of a point in a meridian plane cross-section of the torus obeying the condition $X^2 + Y^2 \leq 1$, and $\varphi$ designates the angular coordinate of this cross-section around the torus. Define a map governing evolution on a single step of discrete time, say, as follows

$$\varphi_{n+1} = 2\varphi_n, \quad X_{n+1} = \tfrac{3}{10} X_n + \tfrac{1}{2} \cos\varphi_n, \quad Y_{n+1} = \tfrac{3}{10} Y_n + \tfrac{1}{2} \sin\varphi_n \ . \qquad (2)$$

Its action is illustrated visually in Figure 1: a set of points filling the torus is transformed to the object of smaller volume in a form of the double loop placed inside the original domain. Also, the result of two-fold application of the map is shown, as well as the attractor itself, which is the solenoid arising as the limit object after the multiple iterations.

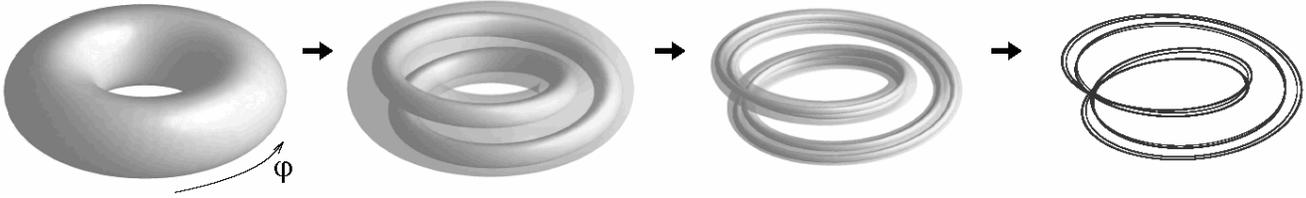

**Figure 1:** Toroidal domain in three-dimensional state space, its transformation after one and two iterations of the map (2), and the Smale-Williams solenoid arising as the limit object after the multiple iterations.

A subtle circumstance is that the geometric construction described above is not equivalent to the map (2) in the topological sense. One can verify this in a home experiment, say, with a real ring-form rubber tube, comparing two versions of the operation. One is that we perform a cut across the tube, stretch in twice in length, fold to get a double loop, and glue the ends, as if the cut was not done at all. It corresponds precisely to the map (2). Another option is a continuous deformation of a circular tube to form a double loop without a cut. It occurs that the last is accompanied by an inevitable additional longitudinal torsion of the tube. Therefore, to achieve the equivalence, one should glue the ends not directly, but with a mutual turn of them by 360°.

If we wish to create a system with continuous time, which gives rise to the solenoid type attractor in the Poincaré map, we have to deal with a continuous deformation of the object in the phase space. This is consistent with the geometric version of the construction; in contrast, the mapping (1) cannot be regarded as an appropriate starting point to reach this purpose.

Recently examples of physically realizable systems with uniformly hyperbolic chaotic attractors were proposed [12-19]. Thereby the preconditions are created for design of actual operating devices, demonstrating structurally stable chaos, for example, in electronics.

Bearing in mind the electronic devices, it is natural to turn to contemporary circuit simulation software products, among which a convenient and popular one is Multisim [20, 21]. Its original version called Electronics Workbench was released in 1995 by the Canadian Interactive Image Technologies Company. Since 2005, improved versions of the software are being developed by National Instruments (in which the original software company came in as a division) under the name of NI Multisim. Results presented in this paper were obtained with the licensed version of the product NI Multisim 10.1.1 purchased by Saratov Branch of IRE RAS.

The first section of the article is devoted to a version of the system proposed in paper [12]. It is composed of two van der Pol oscillators, which become active alternately due to periodic modulation of the parameter responsible for the arising self-oscillations. Both subsystems transfer the excitation to each other in such way that it is accompanied by expanding transformation for the variable characterizing the phases of high-frequency oscillations in successively generated trains. The system is characterized by the presence of attractor of Smale-Williams type in the four-dimensional mapping, which describes the transformation of system states on each period of the modulation. Now, using the Multisim software, the laboratory device is reproduced here, which was studied experimentally in a joint paper of the author with E.P. Seleznev [14].

The second section is devoted to a system built on a base of a single van der Pol oscillator, which by virtue of the periodic parameter modulation manifests alternating stages of activity and depression. Each time at the beginning of the activity stage the excitation is stimulated by a signal emitted during the previous activity stage. The appropriate non-linear transformation provides doubling of the phase of the signal in the delayed feedback circuit. Because of the presence of the delay, the phase space, in which the attractor of the Smale - Williams type is embedded, is infinite dimensional in this case. A scheme in Multisim is considered that corresponds to a laboratory device earlier studied experimentally in collaboration with V.I. Ponomarenko [17].

In the third section, an analog device is proposed, which directly reproduces the mathematical construction of the Smale-Williams solenoid basing on the continuous deformation of a toroidal domain in the three-dimensional state space. The original version of the model was described in [18], but in the present study it has some modifications aimed to simplify the schematic implementation. Note that this system has the minimal dimension of the phase space in which the attractor of Smale-Williams type can occur.

## 1. System with Smale-Williams attractor built on the basis of non-autonomous van der Pol oscillators

In [5] a non-autonomous system was introduced governed by equations

$$\ddot{x} - [A\cos(2\pi t/T) - x^2]\dot{x} + \omega_0^2 x = \varepsilon y \cos\omega_0 t,$$
$$\ddot{y} - [-A\cos(2\pi t/T) - y^2]\dot{y} + 4\omega_0^2 y = \varepsilon x^2.$$
(2)

The variables $x$ and $y$ belong to two van der Pol oscillators excited alternately because of periodic variation of the parameters responsible for the Andronov-Hopf bifurcation. The period, of the modulation is $T$, and its depth is specified by the constant $A$. Operating frequencies of the oscillators are $\omega_0$ and $2\omega_0$. The effect of the subsystems to each other is characterized by coupling parameter $\varepsilon$. The action of the first oscillator onto the second is described by the term proportional to the squared generalized coordinate, and action of the second one onto the first is provided by the term containing product of the generalized coordinate and of the auxiliary signal of frequency $\omega_0$. We assume the relation $T = 2\pi N/\omega_0$ with integer $N$ is valid; so this is a set of equations with periodic coefficients.

Suppose that the first oscillator at some stage of activity is characterized by some phase $\varphi$, i.e. $x \sim \cos(\omega_0 t + \varphi)$. Its action on the second oscillator is determined by the second harmonic $\cos(2\omega_0 t + 2\varphi)$ at the resonance frequency of the second oscillator, which stimulates its excitation at the transition to the active stage; then the resulting oscillations are characterized by the phase $2\varphi$ (up to an additive constant). In turn, the resonant action of the second oscillator on the first is produced due to mixing with the auxiliary signal; it gives rise to appearance of the component at the difference frequency $\omega_0$. It provides stimulation to initiate the oscillations which inherit the phase for the next stage of activity of the first oscillator. Thus, two subsystems in turn transmit the excitation each other, and at successive stages of activity of each oscillator the phase transformations in this approximation are given by the Bernoulli map $\varphi_{n+1} = 2\varphi_n + \text{const } (\text{mod} 2\pi)$.

In accurate description, the map, which describes evolution of the system state on one period of modulation acts in the four-dimensional space $(x, \dot{x}, y, \dot{y})$. It produces stretching for the cyclic variable corresponding to the phase $\varphi$ and compression in the other three directions. One can specify an absorbing domain in the form of a toroid, so that an iteration of the map corresponds to the longitudinal extension and transverse compression of the toroid, and then it is embedded as a double loop inside the original domain; this just corresponds to the construction of the Smale-Williams solenoid [13].[1]

Let us turn to the circuit diagram shown in Figure 2. The scheme is composed of two van der Pol oscillators built on the resonant circuits, one formed by inductor L1 and capacitor C1 and the second, respectively, by L2 and C2. Since the inductances are of identical value, and the capacity in the second subsystem is four times less, the natural frequencies are in the ratio 1:2. With parameters indicated on the diagram, the natural frequencies are approximately 50 kHz and 100 kHz. Negative resistance in each LC-circuit is introduced by a block on the base of an operational amplifier (in the diagram, respectively, U1 and U2). [2] Nonlinear conductivity, which increases the energy losses in the circuits with increase of the amplitude, are introduced by the semiconductor diodes connected in parallel in branches of opposite direction of the current flow (D1, D2 and D3, D4). To modulate the parameter responsible for occurrence of the self-oscillations, each circuit includes a field effect transistor, whose conductivity from the source to the drain is controlled by the voltage on the gate fed from the AC voltage generator at the frequency of 5 kHz. For the parameter modulation in the two oscillators the sources V1 and V2 are used with the opposite polarity of the instant voltages. In one half-cycle of modulation the first oscillator is in the active stage of generation of oscillations, and the second is below the generation threshold. In the next half-cycle, they change roles. The first oscillator effects the second through the multiplier A1, providing the squared signal. The generated second harmonic is used to stimulate excitation of the second oscillator, when his time comes to overcome the generation threshold. In turn, the second oscillator acts on the first through the element A2 mixing the input signal and the auxiliary signal at 50 kHz from the AC voltage source V3. In the output there is a component at difference frequency, which acts on the first oscillator, and stimulates the self-oscillation excitation at the beginning of the next stage of activity.

---

[1] Traditionally, the construction is considered in the three-dimensional space, but here the state space dimension is greater by one then the minimally required.

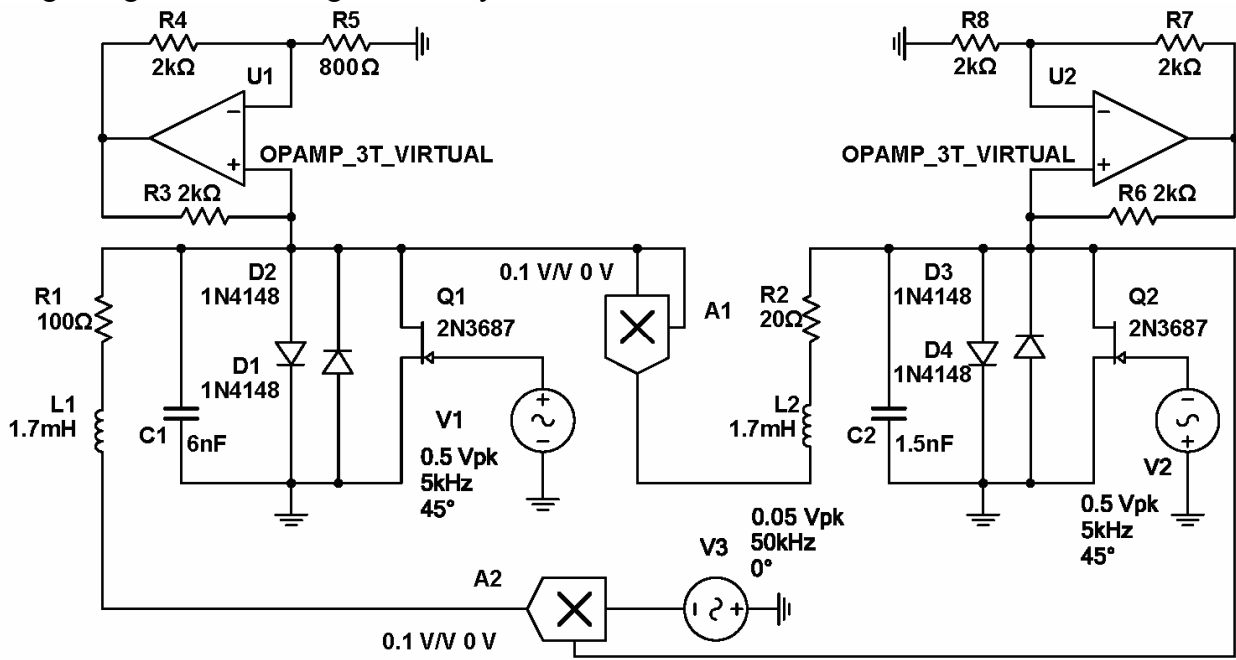

**Figure 2:** Circuit diagram of the device with attractor of Smale – Williams type in the stroboscopic map determining evolution on one modulation period ($T=0.2$ ms). A multiplier A1 has the conversion coefficient equal to 0.1 for transformation from input voltages to output.

Figure 3 shows waveforms for the voltages on the capacitors C1 (upper plot) and C2 (lower plot), obtained in the simulation in Multisim using the Oscilloscope tool (Stimulate - Instruments - Oscilloscope). As seen, in accordance with the above qualitative description, we have the alternate excitation of the oscillations in the contours of L1, C1 and L2, C2.

---

[2] For information relating to circuit design issues, including the functioning and typical application of operational amplifiers, one can recommend books and manuals [22-24]

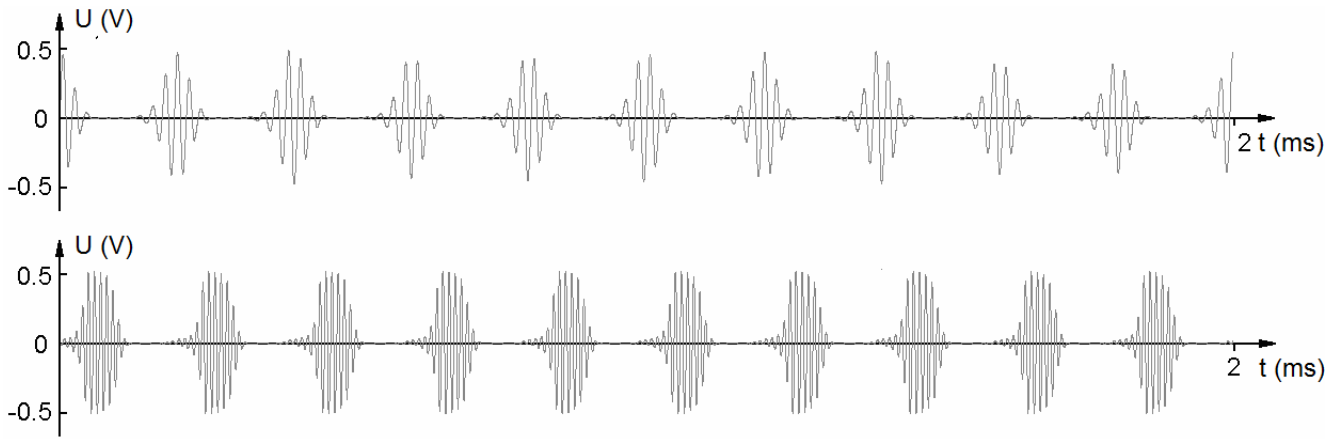

**Figure 3:** Time dependences for the voltages across the capacitor C1 (the upper plot) and C2 (lower plot).

To demonstrate that the transition to each new stage of activity is accompanied by doubling of the variable that characterizes the phase of the oscillations, we proceed as follows. Connect the oscilloscope to have voltage on the capacitor C1 at one input, and the voltage across the resistor R1 at another one. In Multisim the Grapher tool is available, which supports recording data in a file with a possibility of further digital processing. Sampling time step must be set equal to the period of modulation ($T = 0.2$ ms). To choose the right moments for sampling, which correspond to stages of activity of the first oscillator, one can use the phase parameter of the AC sources V1 and V2. (In the diagram, Figure 2, this parameter is assigned equal to 45º, which is appropriate.) When writing in the file, an option Spline Interpolation is preferable among others; this provides better accuracy of the data representation. The file is then processed by an external program, compiled on any user-friendly language (the author used Free Basic IDE). First, both time series, for the voltages on the capacitor and the resistor, $U_C$ and $U_R$, are normalized so that the sums of the squares of the elements for each time series to be of the same value. Then, for each pair of variables $U_C^n, U_R^n$ related to $t=nT$ we define the phase as $\varphi_n = \arg(U_C^n + iU_R^n)$ and relate it to the interval from 0 to $2\pi$. Figure 4 shows a graph which presents the processed data for $\varphi_{n+1}$ versus $\varphi_n$. As seen, the diagram corresponds to the expanding circle map of Bernoulli type. Indeed, one bypass around the circle for the pre-image implies two bypasses for the image, i.e. the condition for the occurrence of the Smale-Williams attractor holds.

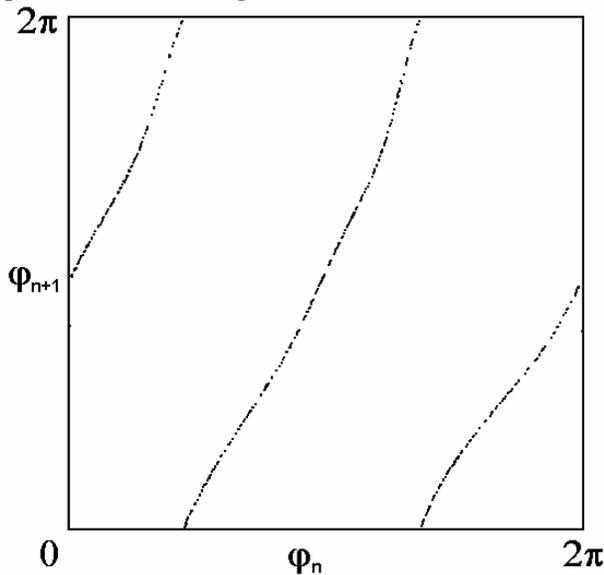

**Figure 4:** An empirical map for the iterative phases.

The same connection of the oscilloscope can be used to produce a portrait of the attractor in the projection on the phase plane of the first oscillator; to do so one has to switch the instrument to the operation mode in which the scan in time is not performed while the horizontal and vertical deflections of the beam are controlled by the input voltages $U_C$ and $U_R$. A portrait of the attractor obtained in such way is shown in Figure 5. To depict the attractor in the stroboscopic section, one can turn again to the recorded data file used in the construction of the iterative phase diagram and present the data graphically in the coordinates ($U_C$, $U_R$) by means of the external program. The diagram is shown in Fig. 5b. Here one can see an object just corresponding to the Smale-Williams solenoid, with its characteristic feature of the transverse Cantor structure.

Figure 6 shows spectra of the signals generated by the two subsystems in the course of operation of the scheme. It is implemented with the Spectrum Analyzer tool in Multisim (Stimulate - Instruments - Spectrum Analyzer) with proper installation of the operating frequency range and resolution of the analysis. The spectra in Figure 6 are given in logarithmic scale: the ordinate is the power spectral density in decibels. Panel (a) corresponds to the spectrum of the signal from the first oscillator (voltage across the capacitor C1), and panel (b) to the signal from the second oscillator (voltage across the capacitor C2). Observed spectra are continuous, as expected for the chaotic signals. For the first oscillator, it concentrates mostly near the natural frequency of 50 kHz, and for the second in the neighborhood of the doubled frequency of 100 kHz. In the low frequency part of the spectra visible discrete components at the frequency of 5 kHz and its harmonics are visible, which appear due to the presence of periodic modulation of parameters at this frequency.

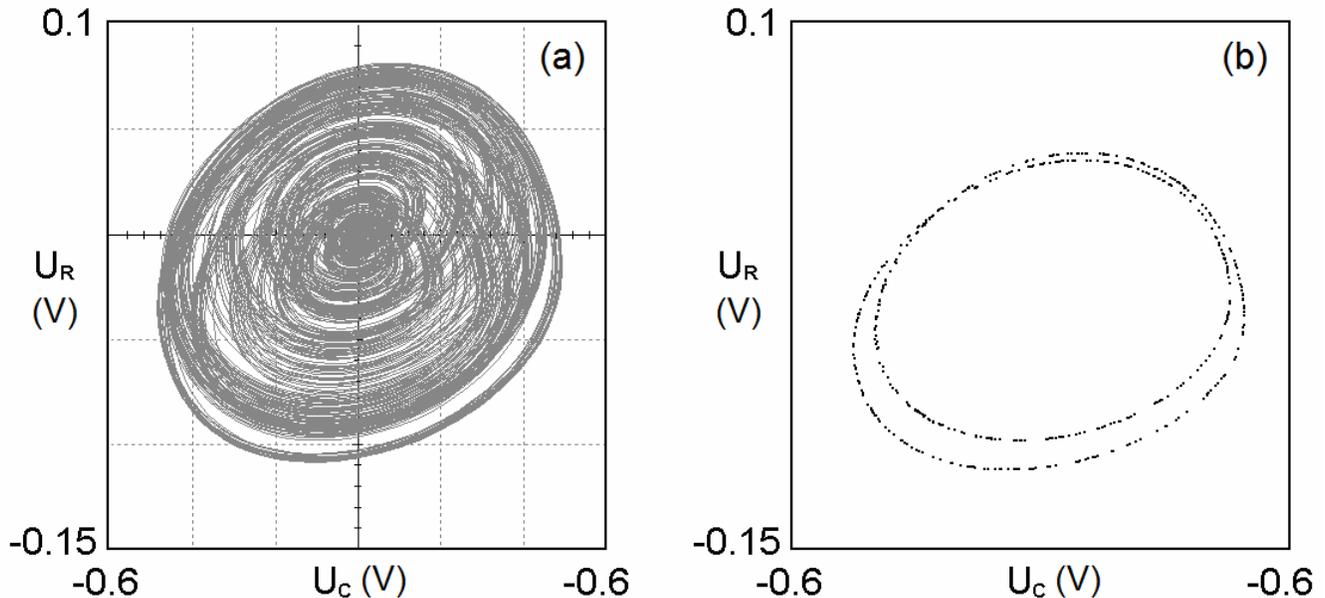

**Figure 5:** Attractor in the projection from the extended phase space (a) and in the stroboscopic section (b). The horizontal and vertical axes correspond, respectively, to the voltages across the capacitor C1 and the resistor R1.

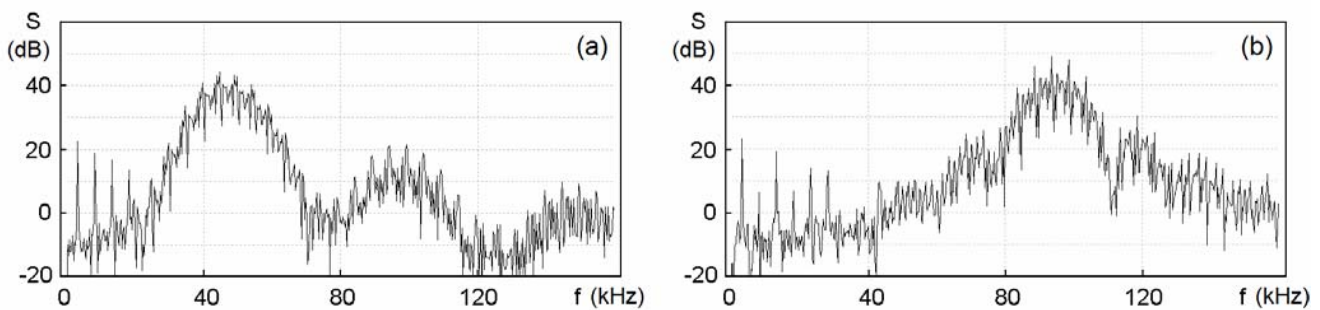

**Figure 6:** Spectra of voltage fluctuations on the capacitor C1 (a) and C2 (b)

## 2. System based on non-autonomous van der Pol oscillator with an additional delayed feedback

An alternative way to implement the expanding circle map for the phases of successive stages of oscillatory activity is based on a use of delayed feedback. In this case, it is sufficient to have a single oscillator, alternately manifesting activity and depression. At each new activity stage the excitation is stimulated by signal from the feedback loop being emitted at the previous stage of activity and having undergone the proper phase transformation. In [18] a system was proposed governed by the delay equation

$$\ddot{x} - [A\cos(2\pi t/T) - x^2]\dot{x} + \omega_0^2 x = \varepsilon x(t-\tau)\dot{x}(t-\tau)\cos\omega_0 t. \qquad (3)$$

Here $x$ is a dynamic variable of the van der Pol oscillator with operating frequency $\omega_0$, in which the parameter responsible for the bifurcation of birth of the limit cycle is slowly modulated in time with period $T$ and amplitude $A$, so that the oscillator is alternately active or damped. In the right-hand part of the equation a term is added responsible for the delayed feedback. This is a product of the dynamical variable at the retarded time, its derivative, and the auxiliary signal of frequency $\omega_0$. The delay time is characterized by the constant $\tau$. Parameter $\varepsilon$ determining the intensity of the feedback is controlled by the amplitude of the auxiliary signal. It is assumed that $N = \omega_0 T/2\pi$ is integer, so that the external driving is periodic in time. Note that the phase space of the system (3) is infinite dimensional: to assign definite instantaneous state one needs not only to indicate the instantaneous coordinate and velocity of the oscillator, but to specify the variable $x$ as a function of time on the interval of duration $\tau$.

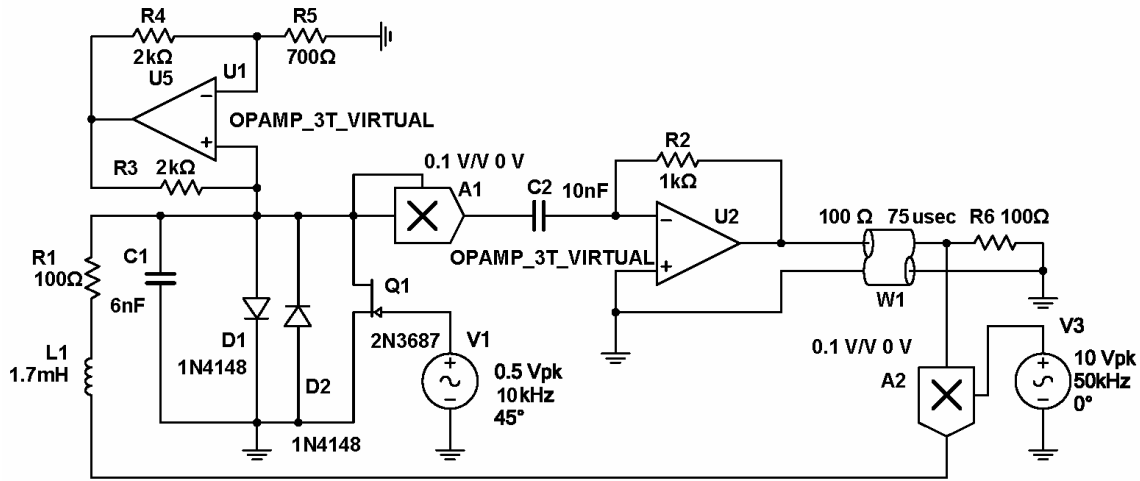

**Figure 7:** Circuit diagram of the device with delayed feedback. Multipliers A1 and A2 have conversion coefficients for transformation from input to output voltages equal to 0.1.

Figure 7 shows circuit diagram of the system. The van der Pol oscillator is implemented with the oscillatory circuit composed of the inductor L1, and the capacitor C1. With parameters indicated on the diagram the natural frequency is about 50 kHz. Negative resistance element is built with the operational amplifier U1. Nonlinear conductivity, which increases the energy loss with amplitude growth, is introduced by semiconductor diodes D1 and D2. To provide parameter modulation the circuit includes the field-effect transistor, whose source-drain conductivity is controlled by the gate alternating voltage with a frequency of 10 kHz from the source V1. In one half-period of the modulation the oscillator is generating, and in the other it is below the excitation threshold. The generated signal is squared by the multiplier A1, transformed through a differentiator built on the capacitor C2, resistor R2 and the operational amplifier U2, and enters the delay line W1. On its output there is a resistor R6 of the value corresponding to the impedance matching with the delay line that provides a lack of reflections. With the multiplier A2 the output signal from the delay line is mixed with the auxiliary signal of frequency 50 kHz from the source V3. The resulting signal is used to stimulate the resonant excitation of the van der Pol oscillator at the beginning of its next stage of activity. The delay time τ=75 ms is selected to be ¾ of the modulation period. This ensures arrival of the signal emitted at the stage of activity through the feedback loop at the right time prior to the onset of the next stage of activity of the oscillator.

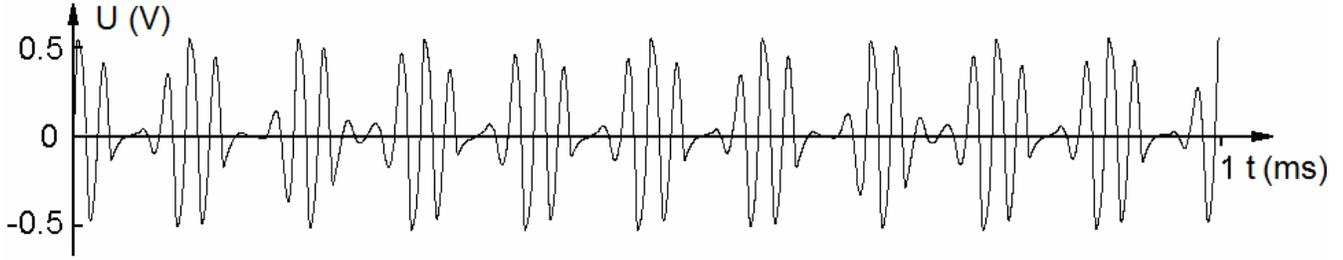

**Figure 8:** Time dependence for the voltage across the capacitor C1.

Figure 8 shows a plot of the voltage on the capacitor C1 as function of time obtained by simulation in Multisim using the Oscilloscope tool. As seen, in the oscillatory circuit L1, C1, in accordance with the above qualitative description, the alternating excitation and damping occur. The phase of the oscillations from one epoch of activity to another varies in accordance with the chaotic expanding circle map. To verify this, connect the oscilloscope so that one input is control voltage on the capacitor C1 and the second is the voltage across the resistor R1, and write data to a file with a sample time step equal to the period of modulation $T$=0.1 ms. The file is then processed by external program in the same way as described in the previous section. Figure 9 shows a graph of $\varphi_{n+1}$ versus $\varphi_n$, which, as observed, corresponds to the expanding circle map of the Bernoulli type.

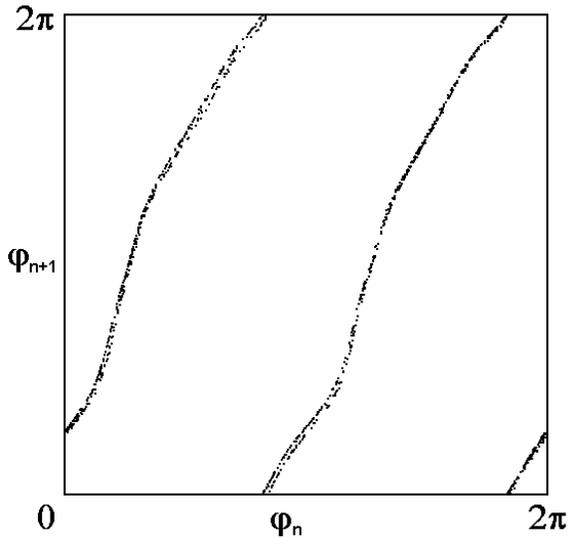

**Figure 9:** An empirical map for the iterative phases.

Figure 10 shows portraits of the attractor in the projection from the infinite-dimensional state space of the system with delay on the plane of the variables $U_C$, $U_R$. The diagram (a) is obtained by switching the oscilloscope into the operation mode, in which the horizontal and vertical deflections of the beam are controlled by the voltages $U_C$ and $U_R$. To depict the attractor in the stroboscopic section, the file is exploited again used in the construction of the phase iterative diagrams; now the data are presented graphically by the external program in the coordinates ($U_C$, $U_R$). The resulting object shown in the panel (b) of Figure 10 corresponds to the solenoid of Smale-Williams type.

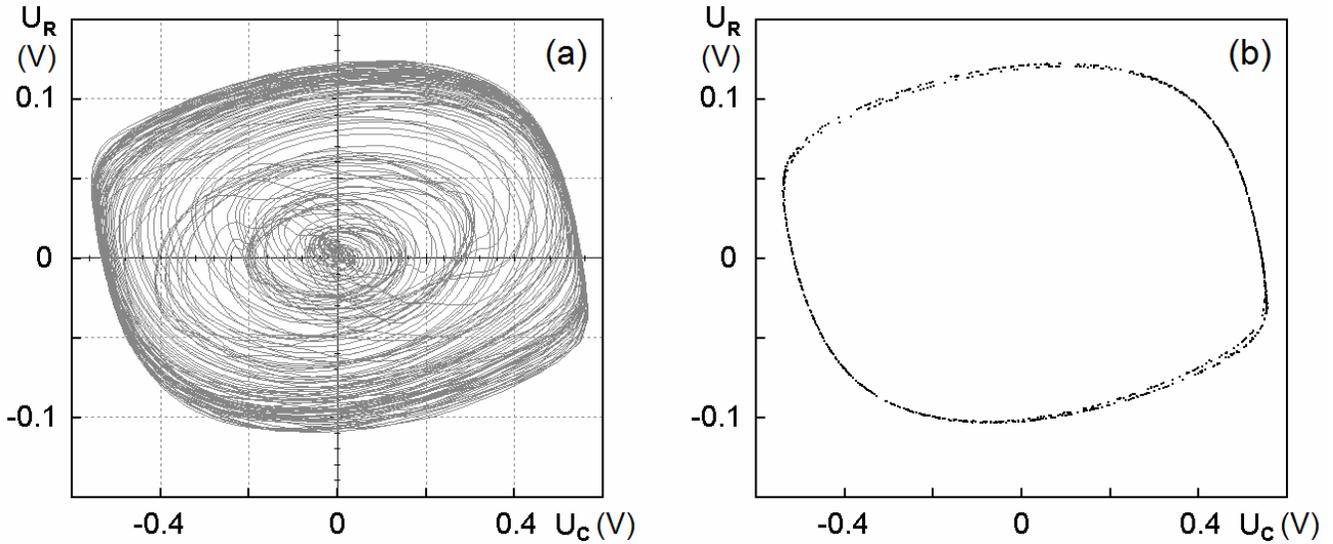

**Figure 10:** Portrait of attractor in projection on the plane from the extended phase space (a) and in the stroboscopic section (b). The horizontal and vertical axes correspond, respectively, to the voltages on the capacitor C1 and the resistor R1.

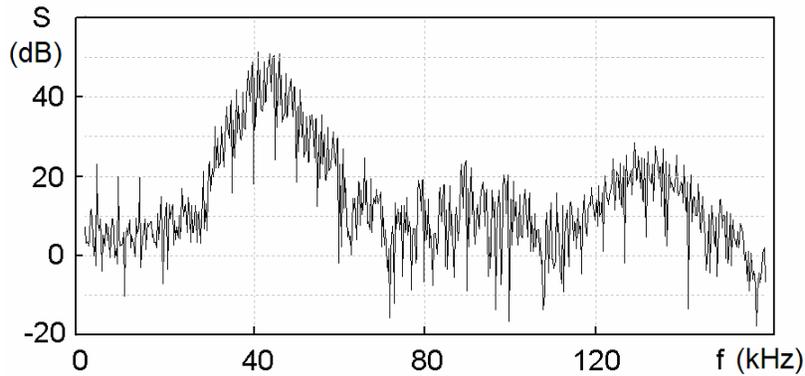

**Figure 11:** The spectrum of oscillating voltage on the capacitor C1.

Figure 11 shows the spectrum built with the Spectrum Analyzer tool in Multisim for the signal representing the voltage across the capacitor C1. Logarithmic scale is used. The spectrum is continuous been concentrated mostly nearby the natural frequency of the van der Pol oscillator (50 kHz). In the low frequency part of the spectrum discrete components are visible at the frequency of 5 kHz and its harmonics, due to the presence of periodic modulation of parameters with that frequency.

## *3. Analog device with Smale-Williams attractor*

Let us construct a non-autonomous system in which the instantaneous state is determined by a 3-dimensional vector $(x, y, z)$, and the extended phase space is of dimension four.

The dynamics in each period of the external driving will consist of three stages. In the first stage the evolution provides attenuation of the variable $z$ to zero while $x$ and $y$ tend to the unit circle in the plane of these variables. Toroidal domain used in the construction of the Smale-Williams attractor corresponds to a neighborhood of this unit circle. In the second stage, in the space of three variables ($x$, $y$, $z$) the differential rotation is applied around the axis of $x$, with an angular velocity linearly depending on $x$ and vanishing at $x=0$. At the edges it must be of such value such that the domain placed initially around the unit circle took a form of figure "eight" in projection on the plane $(x, y)$. At last, during the third stage the plane, near which the figure "eight" is placed, undergoes an inflection, and due to the transversal flattening along the axes $z$ and the shift along the axis $x$, both halves of the figure "eight" take a position that provides embedding of the deformed shape inside the original toroidal domain in the form of a double loop.

The equations describing the dynamics of the system read:

$$\dot{x} = \mu x(1-x^2-y^2) - \tfrac{\pi}{2}\xi_2(t)z,$$
$$\dot{y} = \mu y(1-x^2-y^2) - \tfrac{\pi}{2}\xi_1(t)xz - \xi_2(t)[2y - 3(x^2+z^2)+b], \qquad (4)$$
$$\dot{z} = -\gamma z + \tfrac{\pi}{2}\xi_1(t)xy + \tfrac{\pi}{2}\xi_2(t)x,$$

where $\xi_1(t)$ and $\xi_2(t)$ are periodic functions of time switching on and off the processes responsible for the dynamics on the second and the third stages. Period $T$ is specified to be 10, assuming that the function $\xi_2(t)$ equals to 1 at the end of each period on a unit time interval, and zero for the rest of the period. Function $\xi_1(t)$ equals 1 on the previous unit time interval and zero for the rest of the period. Test calculations show that the assignment of the parameters $\mu=0.3$, $a=1.5$, $b=0.9$, $\gamma=0.2$ provides the desired type of dynamics.

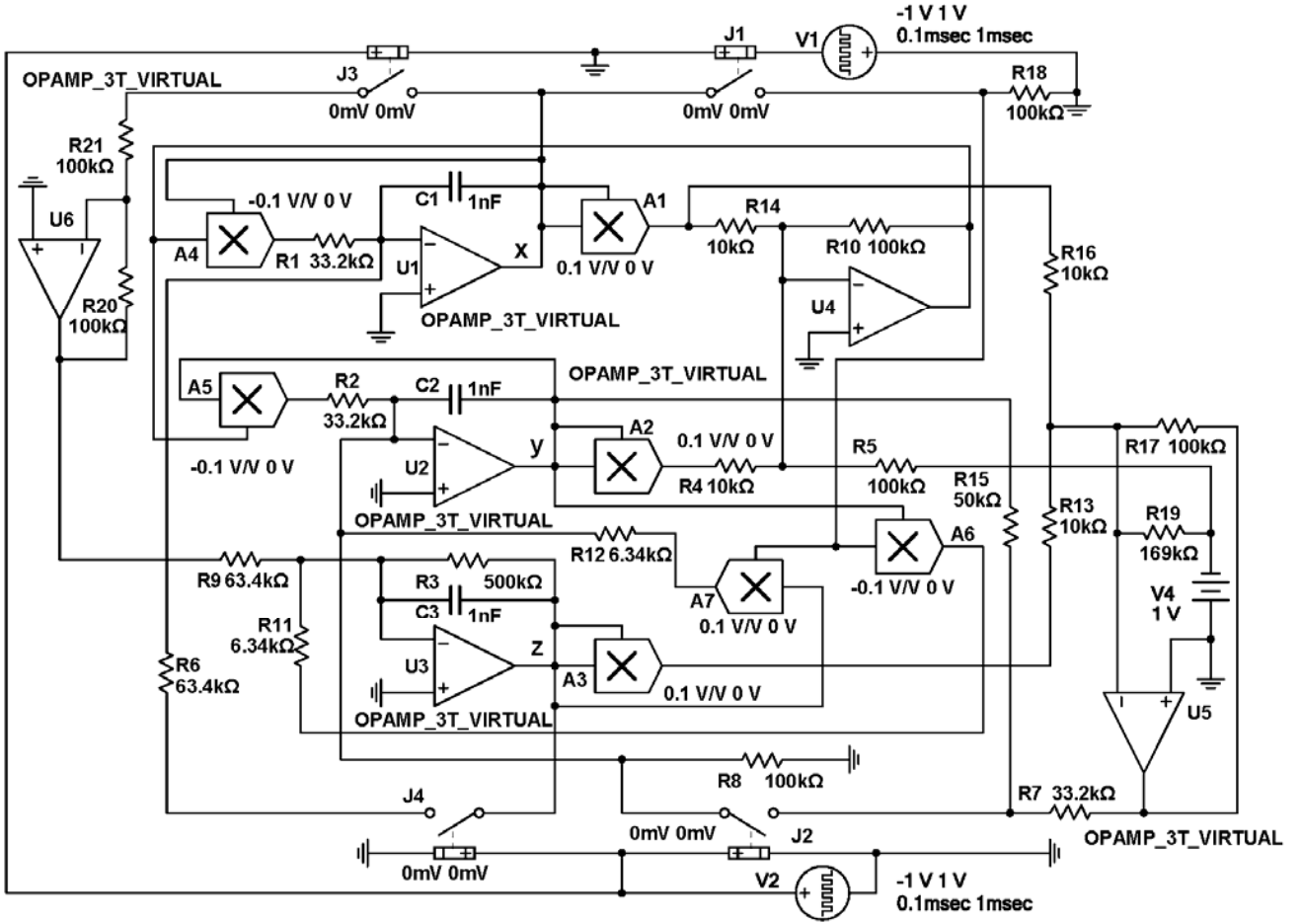

**Figure 12:** Circuit diagram of the analog devices, which provides attractor of Smale-Williams type in the map determining evolution on the period $T=1$ ms. Dynamic variables $x$, $y$, $z$ correspond to the voltages across the capacitors C1, C2 and C3, respectively. Multipliers A1, A2, A3, and A7 have positive conversion coefficients for transformation from input to output voltages, and those for A4, A5, and A6 are negative; the absolute value of the conversion coefficients is 0.1.

Circuit diagram of the device is shown in Figure 12. The variables $x$, $y$, $z$ correspond to output signals of integrators built on the operational amplifiers U1, U2, U3. These are voltages on the capacitors $C_1$, $C_2$, $C_3$ measured in volts. Conventionally, the unit time interval is defined to be $\Delta t = RC = 0.1$ ms, where $C$ is the value of capacitances $C_1$, $C_2$, $C_3$, and the characteristic resistance $R=100\ \Omega$. The period $T$ equals 10 units, i.e. 1 ms. The keys J1, J2, J3, J4 are open or closed with period $T$ by the pulse signals from the sources V1 and V2. In the first stage covering 80% of the total period, all the keys are closed. In the second stage, constituting 10% of the time interval, the key J1 is open, and in the third stage, which also occupies 10% of the period, the keys J2, J3, J4 are open.

By means of the multipliers $A_1$, $A_2$, the inverting summing amplifier U4, and the multipliers $A_4$, $A_5$ the signal is generated providing the evolution of the variables $x$ and $y$ to the situation $x^2 + y^2 = 1$, whereas the presence of resistor R3 provides attenuation of the variable $z$ to zero. When the switch J1 is closed, the signals proportional to $xz$ and $-xy$ (for their formation the multipliers A6 and A7 are used), are fed to integrators associated with the variables $y$ and $z$. This ensures the differential rotation in the phase space around the axis $x$. When the keys J2, J3, J4 are open, the integrators for $x$ and $z$ are fed with signals proportional to, respectively, $z$ and $-x$ that provides the required rotation of the halves of the figure "eight" around the axis $y$, and the integrator $y$ is fed by the signal containing the sum of squares of $x$ and $z$, a constant term, and an added part proportional to $y$.

The indicated in Fig. 12 circuit elements correspond to the above mentioned concrete parameters of the system.

Now, we can demonstrate that the dynamics of the scheme corresponds to the equations (4).

Figure 13 shows the plots for the voltages on the capacitors C1, C2, C3 versus time obtained by simulation using a multi-channel oscilloscope available in the Multisim (Stimulate - Instruments - Four Channel Oscilloscope). Note that the time dependence obviously looks like a non-periodic process.

To reveal the dynamic nature of the observed behavior as associated with attractor of Smale-Williams type the data of the simulation were written to a file in the form of time series with a sampling step equal to the period of modulation $T=1$ ms. Then the time series $x_n$ and $y_n$ are processed by an external program, and for each step $n$ the angular coordinate on the plane $(x, y)$ is determined as $\varphi_n = \arg(x_n + iy_n)$. As seen from Fig. 13, showing $\varphi_{n+1}$ as a function of the previous value $\varphi_n$, each period of modulation is accompanied with transformation of the angular variable according to the expanding circle map of the Bernoulli type. In the presence of compression in other directions in the phase space, this corresponds to the occurrence of attractor of Smale-Williams type. [3]

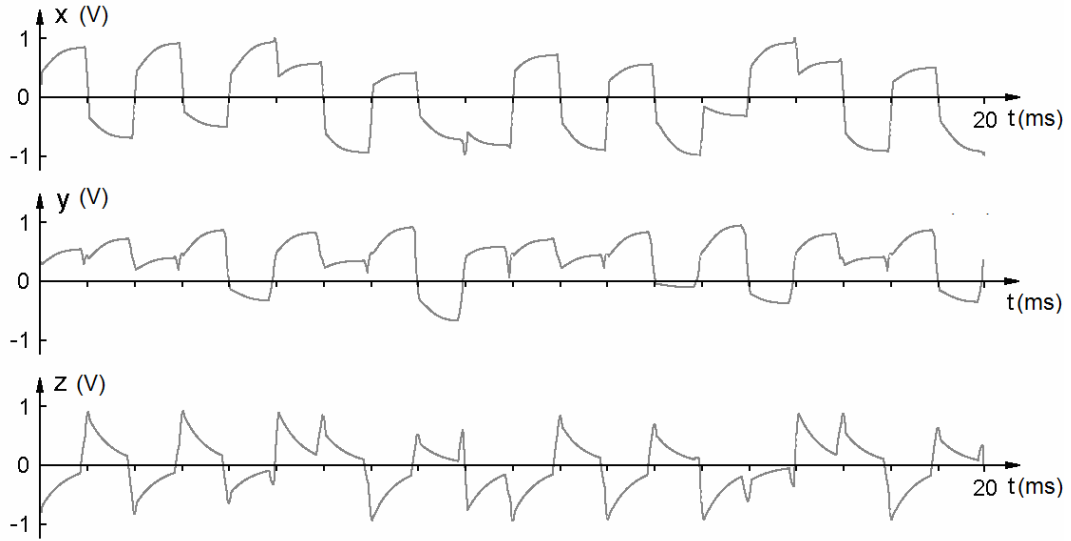

**Figure 13:** Plots of the voltages on the capacitors C1, C2, C3 versus time obtained from simulating the analog devices of Figure 11 in Multisim.

---

[3] Other testing procedures such as evaluation of the spectrum of Lyapunov exponents, check of the transversality of the intersection of manifolds, inspection of the cone criterion etc. are questionable in the approach based exclusively on simulation in Multisim. The respective computations were performed by means of the numerical solution of Equations (4) and respective variation equations by a finite-difference method, and the hyperbolic nature of the attractor was confirmed.

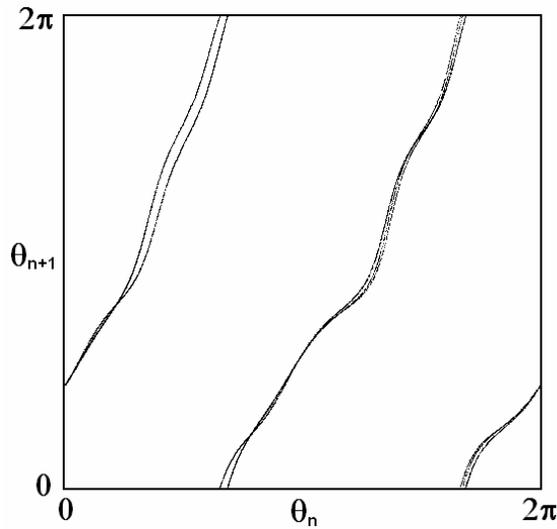

**Figure 14:** Empirical iteration diagram for the angular variable

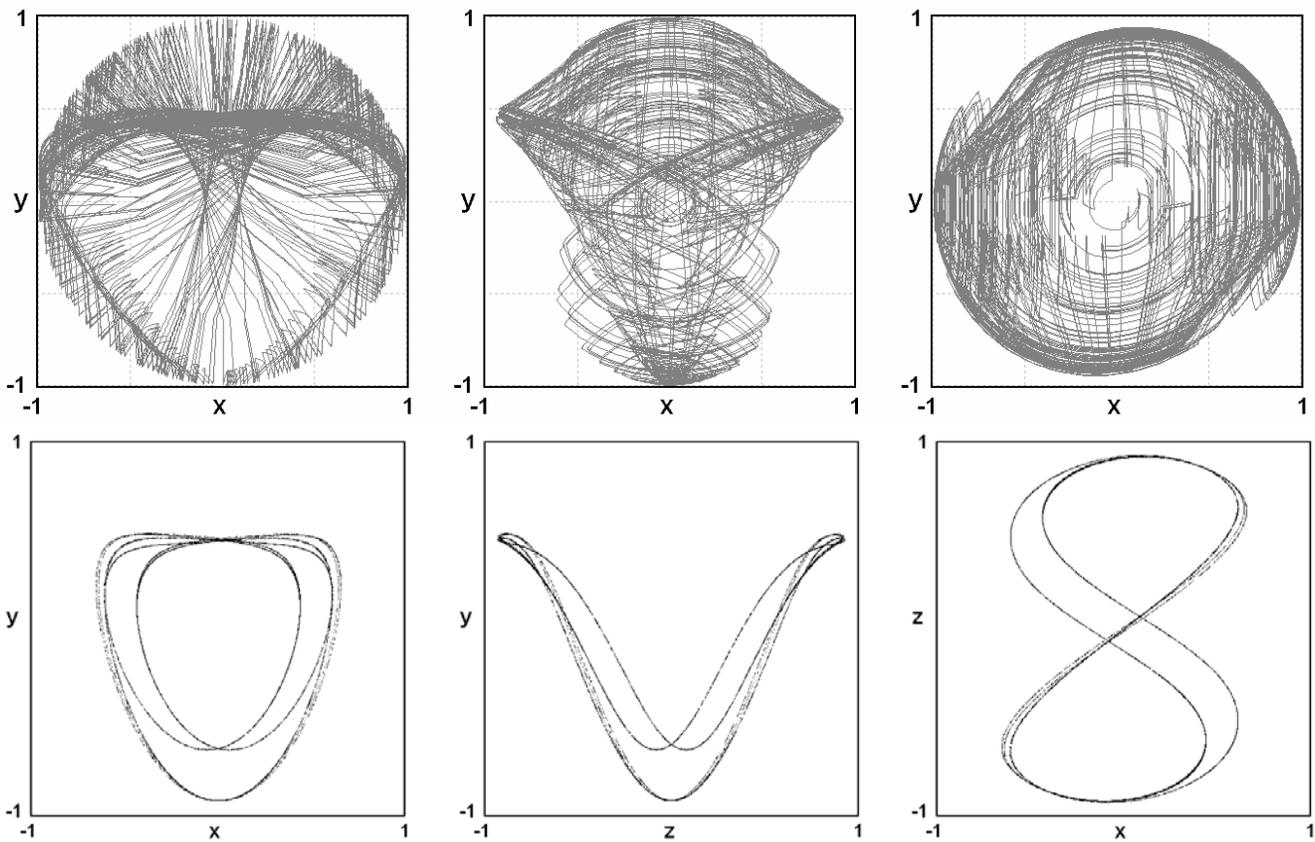

**Figure 15**: Pictures of the attractor in three projections from the extended phase space (a) and the respective portraits in the stroboscopic section (b).

In the top row Figure 15 shows attractor in three projections from the four-dimensional extended phase space, obtained directly in the course of simulation of the dynamics in Multisim with the Oscilloscope tool. To get each image, we connect the oscilloscope to the circuit in such way that the horizontal and vertical deflections of the beam are controlled by the voltages on the capacitors C1 and C2, or C3 and C2, or C1 and C3, respectively, for the first, second and third picture. To construct portraits in the stroboscopic section shown in the bottom row, we use the data file recorded in the course of simulation of the dynamics using multi-channel oscilloscope sampled with time step of one period *T*, and processed for picturing by external computer program. The pictures can be interpreted as images of the Smale-Williams solenoid embedded in the three-dimensional space of states of the stroboscopic map in three projections.

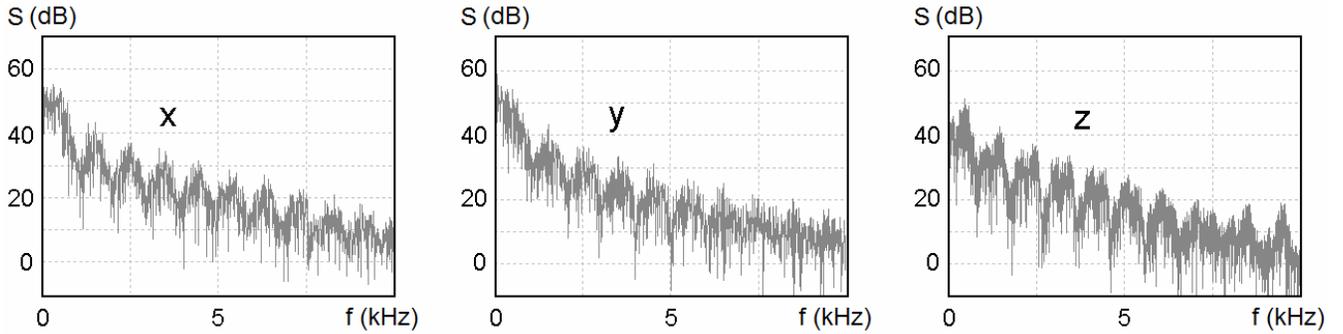

**Figure 16:** Spectra of oscillating voltages on the capacitors C1 (*x*), C2 (*y*) and C3 (*z*).

Figure 16 shows the power spectra recorded by means of the Spectrum Analyzer tool for the signals representing the voltages on the capacitors C1, C2, C3. The plots use the logarithmic scale. The spectra are continuous and occupy a wide frequency range. Irregularity can be noted with a characteristic frequency interval of 1 kHz, which is obviously due to the presence of external modulation of parameters in the system with period $T$=1 ms.

## *4. Conclusion*

The article presents the schemes of electronic devices, which are non-autonomous systems with periodic modulation of parameters manifesting chaotic dynamics associated with attractors of Smale-Williams type in the stroboscopic maps, which describe transformation of the states during the characteristic time periods. Interpretation of attractor as the Smale-Williams solenoid, in each case is based on the fact that we can introduce an angular or phase variable, which for a period of the modulation undergoes the double expanding circle map (the Bernoulli map), whereas in other directions in the state space contraction of the phase volume takes place.

We used two methodologically different approaches, although the line between them is in a sense conventional. Namely, the first two constructions may be regarded as corresponding to elaboration of a physical system that has the type of dynamic behavior similar to that in the model equations. Basis is one or two oscillatory circuits to which elements are added that provide the required type of dynamics (elements of negative resistance, AC and DC sources, feedback elements, etc.). The goal is to promote the desirable type of dynamics at a qualitative level, rather than to reproduce precisely the dynamics corresponding to the original equations. In contrast, in the device considered in section 3 the ideology is adopted of accurate reproduction of the underlying equations exploiting elements traditionally used in analog modeling technology, like integrators, multipliers, adders, etc. Actually, both approaches suggest concrete ways to build real operating electronic devices that act as generators of a structurally stable chaos.

Electronic circuits similar to those described in the paper, may find application in systems of hidden communication [25, 26], noise radar [27], as well as for cryptographic schemes [28,29].

One possible application is generation of random numbers, which is considered as a practically important task [30, 31]. For a system with attractor of the Smale-Williams type, a binary random sequence is obtained in a natural way as a symbolic sequence encoding the trajectory on the attractor:

Zeros or ones simply correspond to visiting one or the other half of the toroidal domain by the angular variable at each next time period in the course of the system operation.[4]

The materials presented here may be proposed for teaching principles of design and analysis of systems with complex dynamics for graduate and postgraduate students specializing in radio-physics and electronics. In particular, we recommend a use of the considered schemes in the computer and experimental laboratory training sessions.

*Paper was prepared with financial support from RFBR (09-02-00426). The author thanks V.I. Ponomarenko and E.P. Seleznev for helpful discussions.*

---

[4] Mathematical models of the considered devices, demonstrating the dynamical chaos may be regarded as the Pseudo-Random Number Generators as it is used to call in literature. However, their implemented as physical devices should be classified as True Random Number Generator, since their operation takes place in the presence of inevitable noise and fluctuations. In fact, in the process of dynamical evolution on the attractor the noises are amplified from the microscopic to macroscopic level by virtue of the inherent sensitivity of chaos to perturbation of the phase trajectories, which occurs due to the presence of a positive Lyapunov exponent. Thus, the systems under the continuous action of a weak noise, select their phase trajectories on the attractor really in a random way.